\documentclass[12pt,a4paper]{article}
\usepackage[latin1]{inputenc}
\usepackage{a4wide}
\usepackage{amssymb, amsmath}
\usepackage{graphicx}

\def\cH{{\cal{H}}}
\def\cD{{\cal{D}}}

\def\cT{{\cal{T}}}
\def\cS{{\cal{S}}}
\def\gR{\mathbb{R}}
\def\gQ{\mathbb{Q}}

\def\gZ{\mathbb{Z}}
\def\gS{\mathbb{S}}

\renewcommand{\theequation}{\thesection.\arabic{equation}}
\begin{document}

\title{Quantum Hall Conductivity in a Landau Type Model with a Realistic Geometry II}
\author{F. Chandelier$^a$, Y. Georgelin$^a$, T. Masson$^b$, J.-C. Wallet$^a$}
\date{}
\maketitle
\begin{center}
$^a$ Groupe de Physique Th\'eorique,\\
Institut de Physique Nucl\'eaire\\
F-91406 Orsay Cedex, France\\
\bigskip
$^b$ Laboratoire de Physique Th\'eorique (UMR 8627)\\
B\^at 210, Universit\'e Paris-Sud Orsay\\
F-91405 Orsay Cedex
\end{center}

\bigskip
\begin{abstract}
We use a mathematical framework that we introduced in a previous paper to study geometrical and quantum mechanical aspects of a Hall system with finite size and general boundary conditions. Geometrical structures control possibly the integral or fractionnal quantization of the Hall conductivity depending on the value of $NB/2\pi$ ($N$ is the number of charge carriers and $B$ is the magnetic field). When $NB/2\pi$ is irrationnal, we show that monovalued wave functions can be constructed only on the graph of a free group with two generators. When $NB/2\pi$ is rationnal, the relevant space becomes a puncturated Riemann surface. We finally discuss our results from a phenomenological viewpoint.
\end{abstract}

\vfill
\begin{flushleft}
LPT-Orsay 03-101\\
IPNO-DR-03-11
\end{flushleft}
\newpage

\section{Introduction}
\setcounter{equation}{0}

It is known that the integral quantization of the Hall conductivity can be understood as resulting from topological features underlying the Quantum Hall systems \cite{PranGirv:90, Ston:92}. The first convincing argument in this direction was given by Laughlin for a system with cylindrical geometry \cite{Laug:81} and was refined in subsequent works considering\footnote{assuming the existence of a spectral gap} the case of non-interacting charges carriers (electrons) in a periodic potential \cite{ThouKohmNighNijs:82, Kohm:85}, then further exented to take into account Quantum Hall systems with disorder and/or interacting electrons \cite{NiuThou:84, NiuThouWu:85, AvroSeil:85, Thou:83}. It was realized that the Kubo formula giving rise to the Hall conductivity can be expressed in term of the integral of the first Chern class of a certain line vector bundle, stemming from the fact that in the quantum mechanical description of these systems two parameters submitted to some periodic conditions can be singled out or introduced. Attempts to explain the observed fractionnal quantization of the Hall conductivity have also appeared in which basically the above (topological) scheme is further combined with additional (physical) plausible assumptions, for instance the existence of a degeneracy of the ground state of the system together with the existence of a gap above it and/or the validity of averaging the Hall conductivity over gauge parameters \cite{NiuThou:84, NiuThouWu:85, AvroSeil:85, Thou:83}.

In most instance, the above topological approach does not permit one to really predict the value for the Hall conductivity starting from basic ingredients pertaining to the quantum mechanocal description of the systems. This feature may reflect the fact that the line vector bundle which shows up at an essential step of the approach does not (in most cases) retain a sufficient amount of the physical properties shared by the real experiments, such as the geometry of the experimental sample, the number of charges carriers $N$, the applied magnetic field $B$\dots Note that many of the quantum mechanical models introduced so far in the topological approach for the Quantum Hall Effect somehow rely on assumptions made on some of the characteristics of the experiment\footnote{Some models consider both the sample and the threads as quantum mechanical objects which is questionnable from an experimental viewpoint.}, such as for instance torus \cite{WenNiu:90} or annular geometry \cite{Halp:82} for the two dimensional sample, which prove convenient to carry on the whole analysis but are sometime questionnable regarding the actual experimental situation. Ultimately, it would be desirable to incorporate into a single framework all the geometrical and physical constraints of the real experiments, control the validity of the hypothesis currently made on the applicability of the Kubo formula (in particular prove the existence of a gap in the Hamiltonian spectrum), study the possible degeneracy of the ground state. This would permit one in particular to determine clearly whether the observed fractionnal quantization of the Hall conductivity can be explainedm as the integral quantization, in terms of topological properties underlying the Hall systems or relies on non topological (additionnal) features.

In a previous paper \cite{CGMW:03}, we considered a Hall system whose geometry, already introduced in \cite{AvroSeil:85, NiuThou:87}, mimics the one for the real experiments, namely a rectangular sample whose edges are connected by wires (see fig.~\ref{fig-general} of section~\ref{previousresults}). This system was assumed to obey general boundary conditions compatible with current and charge conservation and which may account for the respective quantum and classical nature of the electrons inside the sample and those circulating in the wires. We have introduced and discussed a mathematical framework allowing us to treat properly some quantum mechanical aspects specific to the present system with finite size sample and general boundary conditions. These latter in particular force the representation of the considered abstract operator algebra to be reducible. The relevant representation are found to be indexed by two real numbers (defined modulo $2\pi$) forming some ``reciprocical'' space that permits one to obtain from the Hamiltonian the velocity operators appearing in the Kubo formula. The represenation space has been shown, in turn, to be (continuously) decomposable into representaions of a unitary symmetry which connects translations acting \emph{at the same time} on the reciprocical space and on the ``direct'' space built from the physical sample. In \cite{CGMW:03}, we have found that when $NB/2\pi$ takes any integer value, noticeable geometrical structures emerge both in the direct and the reciprocical space: the wave functoins, extended to the whole $\gR^2$-plan, can be interpreted as sections of a line vector bundle over a torus\footnote{This torus in the direct space has no physical meaning and cannot be related in some way to the physical sample.} living in the direct space which is rigidely linked through the above unitary symmetry to a line vector bundle over a torus living in the reciprocical space whose (integral of) first Chern class appears in the Kubo formula. Thanks to this correspondance, this latter formula can be explicitely calculated and we found that the Hall conductivity can take integer and fractionnal values.

When $NB/2\pi$ differs from integer values, the situation becomes far more complicated. It appears that the natural problem to be solved is to determine at least what geometrical structure (if any), that could be used to calculate physical quantities, would survive when $NB/2\pi$ is not integer. This is the purpose of the present paper. The paper is organized as follows. In section~\ref{previousresults}, we summarize the main results obtained in \cite{CGMW:03} and collect the relevant material that will be used in the analysis. The section~\ref{generalcase} is devoted to a characterization of the space on which monovaluated wave functions can be construted for arbitrary (not integer) $NB/2\pi$. When $NB/2\pi$ is irrational, we find that the space is a tree corresponding to the graph of a free group with two generators and can ve identified with the universal covering space of $\gR^2 \setminus \gZ^2$. In section~\ref{rationnalcase}, we consider the case where $NB/2\pi$ takes any irreducible rational value $\ell/k$. Using the van~Kampen theorem, we find that the relevant space on which monovaluated wave functions exist is a Riemann surface with $k^2$ punctures and genus $g_k=1 + k^2(k-1)/2$. The complete proof is rather involved and is presented in the appendix. We further discuss under which circumstances the analysis presented in \cite{CGMW:03} for $NB/2\pi$ integer can be extended to rationnal values of $NB/2\pi$. When the extension can be performed, we find that the Hall conductivity still takes integer or fractionnal values. In section~\ref{physicalcomments}, we discuss from a phenomenological viewpoint a situation when $NB/2\pi$ differs slightly from a rationnal value. The combinations of the fundamental boundary relations for the wave functions with some plausible physical assumptions, in particular the one stating that the geometrical structures exhibited in the pure rationnal case (for which the computation of the Hall conductivity is possible) would survive to the introduction of a sufficiently weak disorder, suggests the existence of an infinite set of domains in the disorder-filling factor plan whose architecture is reminiscent to the global phase diagram from the Quantum Hall Effect proposed in \cite{CGMW:02, CGMW:02b}. Finally we summarize the results and conclude.

\section{$N$-body Landau type Hamiltonian on a square}

\label{previousresults}
\setcounter{equation}{0}

\begin{figure}
\begin{center}
\includegraphics{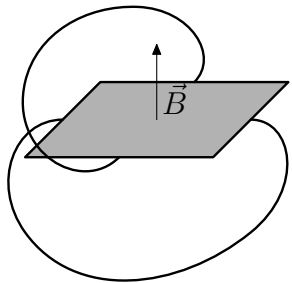}
\end{center}
\caption{Schematic representation of the global geometry of the QHE.}
\label{fig-general}
\end{figure}

In this section, we summarize the results obtained in our previous work \cite{CGMW:03}, hereafter refered as (I), and collect some usefull properties of the corresponding mathematical framework that will be usefull in the sequel. In (I), we considered a spinless multiparticle Landau type Hamiltonian on a finite two-dimensional domain. The Hamiltonian can be expressed in the symmetric gauge\footnote{As shown in (I), the final results do not depend on this gauge choice.} as
\begin{equation}
\label{totalhamiltonian}
H = \sum_{i=1}^N \frac{1}{2 m}\left( (p_{x_i} + \frac{1}{2} B y_i)^2 + (p_{y_i} 
- \frac{1}{2} B x_i)^2\right) + \sum_{1\leq i<j \leq N} V(x_i - x_j, y_i - y_j)
\end{equation}
where $(x_i, y_i)$ (resp. $(p_{x_i}, p_{y_i})$) denote the coordinates (resp. momentum components) of the particle $i$ with mass $m$, $B$ is the external magnetic field applied to the system and $V$ represents a two body interaction potential. The corresponding finite domain was assumed in (I) to be $[0,1]^2 = [0,1] \times [0,1]$, that is, a unit square on the $\gR^2$ plane. Furthermore, the boundary conditions we imposed on this quantum mechanical system were motivated by the geometry of the Quantum Hall Effect, as depicted on fig.~\ref{fig-general}, where \emph{classical} currents circulate in wires connecting apposite edges of the square while no charge accumulation on any edge is possible. As shown in (I), by using the conservation of the currents circulating in the wires together with the absence of charge accumulation on the edges, the computation of the transverse (Hall) conductivity can be performed in extenso from the Kubo formula provided a quantization condition on the product of the number of charge carriers $N$ by the external magnetic field $B$ is satisfied. Namely, one must have $NB=2 \pi \ell$ where $\ell$ is any integer.

Keeping in mind that the physical quantities we are interested in (such as the Hall conductivity) depend only on the collective motion of the $N$ charge carriers, we introduce the following variables
\begin{subequations}
\begin{equation}
\label{collective}
x = \frac{1}{N} \sum_{i=1}^N x_i, \  y = \frac{1}{N} \sum_{i=1}^N y_i, \   p_x = \sum_{i=1}^N p_{x_i}, \  p_y = \sum_{i=1}^N p_{y_i}
\end{equation}
\begin{equation}
\label{internal}
\tilde{x}_i = x_i - x, \quad 
\tilde{y}_i = y_i - y, \quad 
\tilde{p}_{x_i} = p_{x_i} - \frac{1}{N} p_x, \quad  
\tilde{p}_{y_i} = p_{y_i} - \frac{1}{N} p_y
\end{equation}
for $i=1,\dots,N$, where (\ref{collective}) (resp. (\ref{internal})) define the collective (resp. ``internal'') variables and (\ref{internal}) satisfy the relations
\begin{equation}
\sum_{i=1}^N\tilde{x}_i = 0,\quad
\sum_{i=1}^N\tilde{y}_i = 0,\quad
\sum_{i=1}^N\tilde{p}_{x_i} = 0,\quad
\sum_{i=1}^N\tilde{p}_{y_i} = 0.
\end{equation}
\end{subequations}
In these variables, the Hamiltonian (\ref{totalhamiltonian}) can be splitted into two parts as
\begin{subequations}
\begin{align}
H &= H_0 + H_I\\
H_0 &=  \frac{1}{2 N m}\left( (p_{x} + \frac{1}{2} N B y)^2 + (p_{y} - \frac{1}{2} N B x)^2\right)
\end{align}
\end{subequations}
and $H_I$ depends only on the ``internal'' variables. The Hilbert space on which this Hamiltonian acts is itself decomposed into $\cH = \cH_0 \otimes \cH_I$. In the following, we will mainly deal with $H_0$ and $\cH_0$.

The Hilbert space $\cH_0$ looks like $L^2([0,1]^2)$ (for the variables $x$ and $y$) but this identification is not correct as shown in (I). Due to the classical nature of the currents in the wires, the representation of the algebra generated by the operators 
\begin{equation}
P_x = -i\frac{\partial\hfill}{\partial x} + \frac{1}{2} N B y, \ 
P_y = -i\frac{\partial\hfill}{\partial y} - \frac{1}{2} N B x, \ 
H_0 = \frac{1}{2 N m} ( P_x^2 + P_y^2)
\stepcounter{equation}
\label{pxpyh0}
\tag{\theequation a,b,c}
\end{equation}
is reducible. Indeed, in order to define the operators $P_x$ and $P_y$, we have to specify their domain in the Hilbert space. Taking into account the various physical and mathematical constraints, in particular the hypothesis about the geometry of the system depicted on fig.~\ref{fig-general}, the self-adjointness of the operators, we introduce the following domain $\cD_{\gamma, \eta}$:
\begin{align*}
\cD_{\gamma, \eta} = \{ \phi \in L^2([0,1]^2) /\ &
\phi \text{ absolutely continuous},\\
&\frac{\partial \phi}{\partial x}, \frac{\partial \phi}{\partial y} \in L^2([0,1]^2),\\
&\phi(1,y) = e^{i\gamma + \frac{i}{2} NB y} \phi(0,y),\\
&\phi(x,1) = e^{i\eta - \frac{i}{2} NB x} \phi(x,0),\\ 
&\phi(0,0) = \phi(1,0) =\phi(0,1) =\phi(1,1) = 0
\}.
\end{align*}
where $\gamma$ and $\eta$ are two real parameters in the interval $[0, 2\pi[$ which label the possible representations of the above algebra. Notice that in the following we will work with any real values for $\gamma$ and $\eta$, knowing that two values differing by $2\pi$ label the same representation. The arbitrariness in these parameters force us to consider them \emph{all at once}. Then, as discussed in (I), the Hilbert space $\cH_0$ is the direct sum of the Hilbert spaces associated to each individual values of the couple $(\gamma, \eta)$. Because each representation indexed by $(\gamma, \eta)$ takes place in the Hilbert space $L^2([0,1]^2)$, $\cH_0$ is the direct hilbertian integral over $(\gamma, \eta)\in [0, 2\pi[^2$ of $L^2([0,1]^2)$. 

We denote by $(x,y,\gamma, \eta) \mapsto \psi(x,y,\gamma, \eta)$ a general wave function in $\cH_0$ and $\cD$ the common domain of definition of $P_x$ and $P_y$ in $\cH_0$. As shown in (I), any function $\psi \in \cD$ satisfies the boundary conditions
\begin{subequations}
\label{boundarypsi12}
\begin{align}
\label{boundarypsi1}
\psi(1,y,\gamma,\eta) &= e^{i \gamma + \frac{i}{2} NB y} \psi(0,y,\gamma,\eta) \\
\label{boundarypsi2}
\psi(x,1,\gamma,\eta) &= e^{i \eta - \frac{i}{2} NBx} \psi(x,0,\gamma,\eta)
\end{align}
\end{subequations}
These boundary conditions will play a central role in the analysis presented in the next sections, and we will refer to them as the \emph{fundamental boundary relations}. From now on, the $(x,y)$ space will be called the direct space while the $(\gamma, \eta)$ space will be called the reciprocical space. Notice the this terminilogy is somehow inherited from solid state physics since in the present analysis, the variables $(\gamma, \eta)$ look very much like conjugate variables to $(x,y)$. 

Let us now recall the situation in the special case where $NB=2\pi \ell$, where $\ell$ is an integer, hereafter called the integer case. In this situation, the wave functions $\psi \in \cD$ can be extended to the whole plane $(x,y) \in \gR^2$ by the relations
\begin{align}
\label{boundarypsitorus1}
\psi(x+1,y,\gamma,\eta) &= e^{i \gamma + \frac{i}{2} NB y} \psi(x,y,\gamma,\eta) \\
\label{boundarypsitorus2}
\psi(x,y+1,\gamma,\eta) &= e^{i \eta - \frac{i}{2} NB x} \psi(x,y,\gamma,\eta)\ .
\end{align}
Notice that this extension requires explicitely that $NB=2\pi \ell$ is verified, otherwise, an extra phase factor would appear in the wave function when one follows the path $(x,y) \rightarrow (x+1,y) \rightarrow (x+1,y+1) \rightarrow (x,y+1) \rightarrow (x,y)$. Now, for fixed values $(\gamma, \eta$), the extended functions $(x,y) \mapsto \psi(x,y,\gamma,\eta)$ can be interpreted as sections of a line vector bundle over a torus in the variables $(x,y)$ (of periodicity $1$ in $x$ and $y$). This identification was a crucial point in our explicit computation of the Hall conductivity performed in (I). Indeed, our computation made use of the Kubo formula in term of the Chern number of a line vector bundle over the reciprocical space $(\gamma, \eta)$ \cite{Kubo:57, AvroSeilSimo:83} (this space being a torus of periodicity $2\pi$ in our case). Thanks to some symmetries in our system which combine some translations on the two torus \emph{at the same time}, it was possible to caracterise this last line vector bundle over the torus in the variables $(\gamma, \eta)$ using the structure of the line vector bundle over the torus in the variables $(x,y)$. Notice also that in this case, the wave functions $\psi$ can have any values at the point $(0,0)$ and need not to vanish, contrary to the general case. We refer to (I) for details of the computation.

\section{Characterization of the $(x,y)$-direct space for arbitrary $NB$}

\label{generalcase}
\setcounter{equation}{0}

In this section, we study some features of the direct space when $NB$ takes an arbitrary value. As we will see, the situation is far more complicated than the ``integer'' case $NB = 2 \pi \ell$. Our starting point are the fundamental boundary relations (\ref{boundarypsi1}) and (\ref{boundarypsi2}) which caracterize the domain $\cD \subset \cH_0$.

As in the integer case, we want to extend the wave functions on the whole plane $(x,y) \in \gR^2$. Unfortunately, this is not possible as we will see. Let $\psi \in \cD$, a wave function. First, using the boundary relation (\ref{boundarypsi1}), we extend $\psi$ in the $x$ direction by defining ($H$ stands for ``horizontal'')
\begin{equation}
\label{definitionPsiH}
\psi_H(x+p, y, \gamma, \eta) = e^{i p \gamma + \frac{i}{2} p N B y} \psi(x, y, \gamma, \eta)
\end{equation}
for any integer $p$ and any $x \in [0,1[$. The function $\psi_H$ is then defined for $(x,y) \in \gR \times [0,1]$, and satisfies to the boundary relation in $y$
\begin{equation}
\label{boundarypsiHy}
\psi_H(x+p, 1, \gamma, \eta) = e^{i \eta + i p NB - \frac{i}{2} N B (x+p)} \psi_H(x+p, 0, \gamma, \eta)
\end{equation}
This relation is somehow similar to the fundamental boundary relation (\ref{boundarypsi2}) when $x$ is replaced by $x+p$, up to the presence of the extra phase $p NB$.

Now, using the boundary relation (\ref{boundarypsi2}), $\psi$ can be extended in the $y$-direction by defining ($V$ stands for ``vertical'')
\begin{equation}
\label{definitionPsiV}
\psi_V(x, y+q, \gamma, \eta) = e^{i q \eta - \frac{i}{2} q N B x} \psi(x, y, \gamma, \eta)
\end{equation}
for any integer $q$ and any $y \in [0,1[$. The function $\psi_V$ is then defined for $(x,y) \in [0,1] \times \gR$ and satisfies
\begin{equation}
\label{boundarypsiVx}
\psi_V(1, y+q, \gamma, \eta) = e^{i \gamma - i q NB + \frac{i}{2} N B (y+q)} \psi_V(0, y+q, \gamma, \eta)
\end{equation}
Because $\psi_H$ and $\psi_V$ coincide on the square $(x,y) \in [0,1] \times [0,1]$ with $\psi$ (recall that $\psi$ vanishes on the four corners of this square), the function $\psi$ has been extended to the domain $(x,y) \in \gR \times [0,1] \cup [0,1] \times \gR$. 

Let us consider now $\psi_H$ on the square $(x,y) \in [1,2] \times [0,1]$. Using the boundary relation (\ref{boundarypsiHy}), we can extend it to $(x,y)\in [1,2] \times \gR$ through
\begin{equation}
\label{definitionPsiHV}
\psi_{HV}(x, y+q, \gamma, \eta) = e^{i q \eta + i q NB - \frac{i}{2} q N B x} \psi_H(x, y, \gamma, \eta)
\end{equation}
for any integer $q$, any $y \in [0,1[$ and any $x \in [1,2]$. Performing the same procedure, with $\psi_V$ combined with the relation (\ref{boundarypsiVx}), we define $\psi_{VH}$ on $(x,y) \in \gR \times [1,2]$ by
\begin{equation}
\label{definitionPsiVH}
\psi_{VH}(x+p, y, \gamma, \eta) = e^{i p \gamma - i p NB + \frac{i}{2} p N B y} \psi_V(x, y, \gamma, \eta)
\end{equation}
The two functions $\psi_{HV}$ and $\psi_{VH}$ are defined on the common domain $(x,y) \in [1,2] \times [1,2]$. Unfortunately, they do not necessary coincide there! Indeed, for $(x,y) \in [0,1] \times [0,1]$, a straighforward computation using the above definitions yields
\begin{subequations}
\begin{align}
\psi_{HV}(x+1, y+1, \gamma, \eta) &= e^{i \eta + i NB - \frac{i}{2} NB (x+1)} e^{i \gamma + \frac{i}{2} NB y} \psi(x, y, \gamma, \eta) \\
\psi_{VH}(x+1, y+1, \gamma, \eta) &=  e^{i \gamma - i NB + \frac{i}{2} NB (y+1)} e^{i \eta - \frac{i}{2} NB x} \psi(x, y, \gamma, \eta)
\end{align}
\end{subequations}
Thus, we have
\begin{equation}
\label{phase1}
\psi_{HV}(x+1, y+1, \gamma, \eta) = e^{i NB} \psi_{VH}(x+1, y+1, \gamma, \eta)
\end{equation}
When $NB$ is not a multiple of $2\pi$, these two functions differ from each other by an extra phase factor. This therefore leads to \emph{two} different extensions of $\psi$ on the square $[1,2] \times [1,2]$. Notice that the choice of the square $[1,2] \times [1,2]$ is arbitrary in this conclusion. Finally, a straighforward computation shows that $\psi_{HV}$ and $\psi_{VH}$ satisfy the boundary relations
\begin{subequations}
\begin{align}
\label{boundarypsiHVx}
\psi_{HV}(2, y+q, \gamma, \eta) &= e^{i \gamma - i q NB + \frac{i}{2} N B (y+q)} \psi_{HV}(1, y+q, \gamma, \eta)\\
\label{boundarypsiVHy}
\psi_{VH}(x+p, 2, \gamma, \eta) &= e^{i \eta + i p NB - \frac{i}{2} N B (x+p)} \psi_{VH}(x+p, 1, \gamma, \eta)
\end{align}
\end{subequations}
which are equivalent to (\ref{boundarypsiVx}) and (\ref{boundarypsiHy}) respectively.

Let us now generalize this construction. Let $\chi_H$ be a given extension of $\psi$ on a domain $(x,y) \in \gR \times [n,n+1]$ for an integer $n$, with some boundary relations of the form (\ref{boundarypsiHy}) relating $y=n$ to $y=n+1$. It is easy to define a new extension $\chi_{HV}$ on a domain of the form $[m, m+1] \times \gR$ using the common subspace $[m, m+1] \times [n,n+1]$ and a relation generalizing (\ref{definitionPsiHV}). In a similar way, it is possible to extend a function $\chi_V$ defined on a domain $[m, m+1] \times \gR$, with boundary relations in the $x$ direction similar to (\ref{boundarypsiVx}), into a function $\chi_{VH}$ defined on a domain $\gR \times [n,n+1]$, using some natural generalisation of (\ref{definitionPsiVH}).

\begin{figure}
\begin{center}
\includegraphics{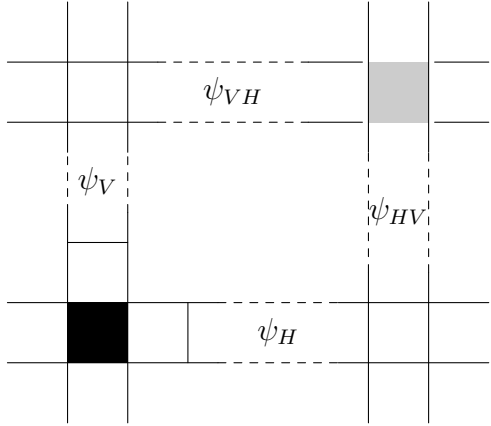}
\end{center}
\caption{The ``fundamental'' domain is the black square. The wave function $\psi$ can be extended on the four stripes as $\psi_{H}$, $\psi_{V}$, $\psi_{HV}$ and $\psi_{VH}$, but on the grey square, $\psi_{HV}$ and $\psi_{VH}$ do not necessary coincide. The phase difference between them is of the form $A \times NB$ where $A$ is the area (counted in number of squares) enclosed by the stripes.}
\label{fig-path}
\end{figure}

Then, starting from the ``fundamental'' square $[0,1] \times [0,1]$, it is possible to define an extension of $\psi$ on any square $[m,m+1] \times [n,n+1]$ in the plane. Unfortunately, the value of this extension \emph{depends} on the path used to go from the fundamental square to the final square. For instance, on fig.~\ref{fig-path}, two extensions in two steps are possible from the fundamental square (in black) to the target square (in grey): a first extension $\psi_{HV}$, for which the first step in an horizontal extension ($\psi_H$) and the second step is vertical, and a second extension, $\psi_{VH}$, for which the first step is vertical ($\psi_V$) and the second horizontal. On the grey square, there is a phase difference between $\psi_{HV}$ and $\psi_{VH}$. Some elementary computation shows that this phase difference is of the form
\begin{equation}
A \times NB
\end{equation}
where $A$ is the area (evaluated in term of the number of squares) limited by the interior of the two paths joining the black square to the grey square used to define the two extensions $\psi_{HV}$ and $\psi_{VH}$. In (\ref{phase1}), this area is just $1$. Note that if the grey square is located at $[m,m+1] \times [n,n+1]$ in $\gR^2$, one has $A = m\times n$.

This ``area rule'' can be easily generalized to any path used to extend $\psi$ from the fundamental square to any target square. This rule will play a very important role in the following.

\begin{figure}
\begin{center}
\includegraphics{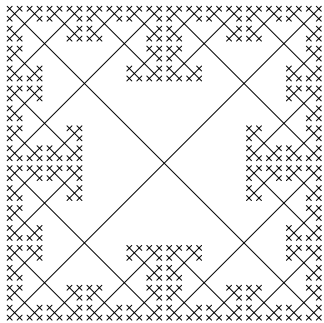}
\end{center}
\caption{This tree, denoted by $\cT$, represents the direct space one can use to define a monovalued extension of the wave function $\psi \in \cD$.}
\label{fig-tree}
\end{figure}

Since the extensions are multivalued, the space on which we want to introduce them cannot be $\gR^2$. The correct space on which an unique extension of $\psi$ can be defined is represented in fig.~\ref{fig-tree}. In this (infinite) tree representation, the squares have been shrunk to vertices, and the links connecting these vertices indicate that two squares are adjacent. 

This tree is constructed from the fundamental square (with removed corners), representented by the vertex at the center, by connecting four squares (with removed corners) at its four edges, and then by iterating this proceduce on any new introduced square, with the important rule that none of these squares are identified with any other, even if they have the same location in $\gR^2$. In the representation of fig.~\ref{fig-tree}, at each step, a scale factor has been applied to avoid some overlapping. If one tries to represent this construction with squares and without any scale factor, one need at least a 3-dimensionnal representation, in which each square in $\gR^2$ is replaced by a stacking of an infinite number of copies. Even the fundamental square is replaced by an infinite number of copies. Mathematically, this tree corresponds to the graph of a two generators free group. One of these generators acts by translation in one direction, and the other as a translation in the orthogonal direction. This space is also the universal covering space of $\gR^2 \setminus \gZ^2$, the space $\gR^2$ where all the points with integer coordinates has been removed. We will denote by $\cT$ this covering space. Then we have a projection map $p : \cT \rightarrow \gR^2 \setminus \gZ^2$, and the inverse image of any square (without corners) $p^{-1}([m,m+1]\times [n,n+1])$ is an infinite number of squares, which we will call the ``covering squares'' of $[m,m+1]\times [n,n+1]$. This last caracterization makes $\cT$ the right space to extend the wave functions $\psi \in \cD$ into a wave function $\widetilde{\psi}$ on $\cT$. Indeed, for any $\psi \in \cD$, we have introduced a procedure to extend $\psi$ on any square in $\gR^2 \setminus \gZ^2$. This extension depends on the path used to build $\psi$  step by step starting from the fundamental square and going to the final square. Because in the universal covering space of $\gR^2 \setminus \gZ^2$ two such paths never meet, we can extend $\psi$ without ambiguities into a monovaluated wave function $\widetilde{\psi}$ on $\cT$. Notice that each path connecting two squares can be decomposed into a ordered family of elementary translation on $\cT$ (which goes from one square to one of its adjacent squares). These elementary translations are representation of the two generators of a free group. This explain why $\cT$ is related to such a group.

When $NB$ is of the form $2 \pi r$ for an \emph{irrationnal} $r$ (which we call the ``irrationnal case'' in the following), the area rule tells us that the phase difference for the values of the extended wave function on two squares covering a square in $\gR^2 \setminus \gZ^2$ is never a multiple of $2 \pi$. This means that the values of $\widetilde{\psi}$ on the covering squares in any stack are always different (except if $\widetilde{\psi}$ vanishes there). On the contrary, when $NB$ is of the form $2 \pi \frac{\ell}{k}$, for any rationnal $\frac{\ell}{k}$ (the ``rationnal case''), the phase difference can be a multiple of $2 \pi$ if the area is a multiple of $k$. It is desirable to identify two squares when the values of the extension of $\psi$ coincide on them. This is the purpose of the next section.

\section{The rationnal case}

\label{rationnalcase}
\setcounter{equation}{0}

In this section, we assume that $NB=2 \pi \frac{\ell}{k}$ where $\frac{\ell}{k}$ is an irreducible rationnal number. When $NB=2 \pi \frac{\ell}{k}$, it appears that the relevant space on which a monovalued extension of the wave function $\psi$ is defined is smaller than $\cT$. This space is obtained from $\cT$ by identifying some covering squares by using a simple rule that will be explained in a while. This space is a compact Riemann surface on which some points are removed (punctures), and depends only on the integer $k$. We turn now on the complete characterisation of these surfaces.

Recall that the main motivation for introducing $\cT$ was to obtain a monovaluated extension $\widetilde{\psi}$ of $\psi \in \cD$. As already noticed in section~\ref{generalcase}, the space $\cT$ is well suited to deal with the irrational case and cannot be reduced to a smaller one. In the present case $NB=2 \pi \frac{\ell}{k}$, the situation is somehow different because the values of the extension of $\psi$ on differents covering squares over the same square in  $\gR^2 \setminus \gZ^2$ can be equal. For instance, the values of the functions $\psi_{HV}$ and $\psi_{VH}$ in fig.~\ref{fig-path} coincide when $mn$ is a multiple of $k$.

We identify two covering squares over the same square in $\gR^2 \setminus \gZ^2$ when the values of $\widetilde{\psi}$ are equal on them (for any $\psi \in \cD$). Then, using the area rule, it is easy to show that the number of covering squares after this identification has been performed is exactly $k$. We denote by $\cT_k$ the corresponding space which is a $k$-fold covering of $\gR^2 \setminus \gZ^2$, and is no longer simply connected.

\begin{figure}
\begin{center}
\includegraphics{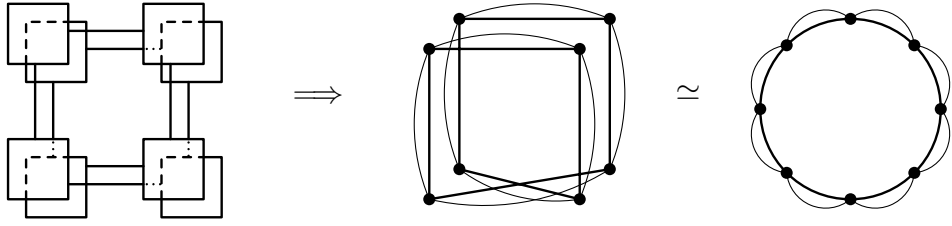}
\end{center}
\caption{Case $k=2$: a $2\times 2$ extract of $\cT_2$ on the left, $\cS_2$ in the middle and on the right.}
\label{fig-k2}
\end{figure}

Let us illustrate the situation in the simplest case $k=2$, as shown in the left graphic of fig.~\ref{fig-k2}. There, four squares in $\gR^2 \setminus \gZ^2$ are considered. Each one has two covering squares, represented by the four stacks of two squares. A link connecting two squares means that these two squares are adjacents. Notice that starting with the top-upper-left square, and following the links, we have to go round \emph{twice} in order to come back, which illustrates the area rule. Obviously, $\cT_2$ is composed of an infinity of such stacks of two squares, and the links which connect these squares. 

Now, on $\cT_k$, it is easy to show that the extended wave function $\widetilde{\psi}$ satisfy the boundary relations
\begin{subequations}
\begin{align}
\label{boundarytildepsix}
\widetilde{\psi}(x+k, y, \gamma, \eta) &= e^{i k \gamma + \frac{i}{2} 2 \pi \ell y} \widetilde{\psi}(x, y, \gamma, \eta)\\
\label{boundarytildepsiy}
\widetilde{\psi}(x, y+k, \gamma, \eta) &= e^{i k \eta - \frac{i}{2} 2 \pi \ell x} \widetilde{\psi}(x, y, \gamma, \eta)
\end{align}
\end{subequations}
because the phases ``$p NB$'' or ``$q NB$'' (see for instance formula (\ref{boundarypsiVHy}) and (\ref{boundarypsiHVx})) are exactly $2\pi \ell$ when $p=q=k$ and then can be eliminated. With these boundary relations, it is possible to interpret $\widetilde{\psi}$ as a section of a line vector bundle over a manifold that we describe now. These boundary relations reveal a periodicity in the variable $x$ and $y$ of length $k$. This means that $\cT_k$ can be reduced by identifying the extreme edges of any stripe of length $k$ ($k$ adjacent squares in the same direction). We denote by $\cS_k$ the manifold obtained from $\cT_k$ by these identifications. 

For instance, for $k=2$, this means that the left edge of the upper-left square is identified with the right edge of the upper-right square (the two squares being on top of their stack). Performing this identification for any square, one obtains the middle graphic of fig.~\ref{fig-k2}, where the squares have been shrunk to some points, and the curved links represent these last identifications. The last graphic on the left represents an unfold version, which is topologically equivalent to $\cS_2$.

\begin{figure}
\begin{center}
\includegraphics{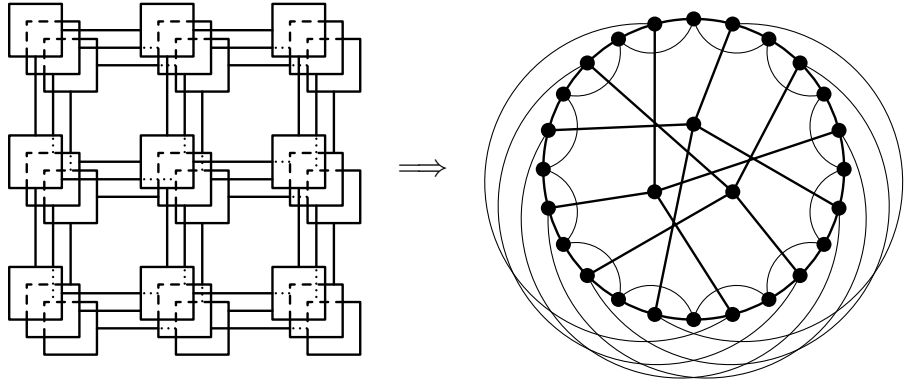}
\end{center}
\caption{Case $k=3$: a $3\times 3$ extract of $\cT_3$ on the left, $\cS_3$ on the right.}
\label{fig-k3}
\end{figure}

For $k=3$, it is still possible to represent graphically these steps, as done in fig.~\ref{fig-k3}: a $3\times 3$ extract of $\cT_3$ on the left and $\cS_3$ on the right. Notice that on $\cT_3$, if one follows some links from square to square, one need to enclose an area which must be a multiple of 3 to come back at the beginning. This is an illustration of the area rule. 

As it can be suspected on fig.~\ref{fig-k3}, when $k$ takes higher values, it is very difficult to represent $\cS_k$. Nevertheless, it is possible to completely caracterize the topology of $\cS_k$ for any $k$. 

Let us give some general properties of $\cS_k$ that can be obtained directly from its construction. First $\cT$ is the universal covering of $\cS_k$, and $\cT_k$ is an intermediaire covering. The two dimensionnal manifold (an orientable finite type Riemann surface \cite{FarKra:80}) $\cS_k$ is constructed by glueing together $k^3$ ``elementary'' squares, the final surface does not have boundary, but it has $k^2$ punctures. These punctures come from the removed points of $\gR^2 \setminus \gZ^2$. Beside the $k$-fold covering of $\cT_k$, each point in $\gZ^2$ is removed only once because in $\cT_k$ one can circulate once around such a point by following the links connecting the squares for which this point is a corner. This can be verified on fig.~\ref{fig-k2} or fig.~\ref{fig-k3}. This is just a consequence of the area rule.

For $k=1$, this construction can be done very easily, and $\cS_1$ is just the ordinary torus with one puncture. This case corresponds to the study presented in (I) when $NB=2 \pi \ell$. In that study, the puncture was removable (because $k=1$), and the entire torus was our base space for the line vector bundle for which the wave function was a section. For $k=2$, the glueing of the $8$ elementary squares is also possible, and gives rise to the Riemann surface of genus $3$ (a sphere $\gS^2$ with $3$ handles) with $4$ punctures. The case $k=3$ is also tractable, and produces the Riemann surface of genus $10$ with $9$ punctures.

We give in Appendix~\ref{pi1} a proof of the following result:
\begin{center}
\emph{$\cS_k$ is a Riemann surface of genus $g_k=\frac{1}{2}(2 + k^2(k-1))$ with $k^2$ punctures.}
\end{center}

Let us comment the results obtained in this section. When $NB=2\pi \frac{\ell}{k}$, the wave functions have a natural interpretation in terms of sections of a line vector bundle over a Riemann surface with $k^2$ puncture $\cS_k$. When $k=1$, which is that case studied in (I), $\cS_1$ is a torus with one puncture and the explicit computation of the transverse conductivity $\sigma_{xy}$ using a geometrical interpretation of the Kubo formula was possible. The extension to the general situation where $\cS_k$, $k>1$, comes into play, becomes extremely more complicated even for $k=2$, so that a direct computation based on the Kubo formula would be untractable. However, this technical difficulty can be partially overcome as we will see in a while.

As explained above, the Riemann surface $\cS_k$ is constructed by gluing $k^3$ copies of the ``fundamental square'' which is nothing but the real physical sample involved in the experiment submitted to the external magnetic field. However, $\cS_k$ is just a mathematical tool which indicates that the relevant wave functions, as sections of some line vector bundle, must satisfy some phase coherence on a distance larger that the length of the real physical sample (because $\cS_k$ is constructed from $k$ copies of the fundamental square in the $x$ and $y$ directions). This last observation will be examined more closely in the next section.

The total magnetic flux through the fundamental square is $NB=2\pi \frac{\ell}{k}$ (as seen by the center of mass of the system). This means that $\cS_k$ is submitted a flux $\Phi = NB\times k^3= 2 \pi \ell k^2$. If $\cS_k$ were closed (without punctures), this outgoing flux could only be produced by a monopole inside the Riemann surface. But thanks to the presence of $k^2$ punctures, this flux can penetrate into the surface, each puncture carrying a penetrating flux of $2 \pi \ell k^2/k^2= 2 \pi \ell$.

Due to the complexity of $\cS_k$, the direct computation of the transverse conductivity from a Kubo formula cannot be performed so that one has to evaluate it with an other method. Using simple arguments and natural physical hypothesis, it is possible to rely the computation of $\sigma_{xy}$ to the result we have obtained in (I). Indeed, the relation $NB=2\pi \frac{\ell}{k}$ can be written as $kNB=2 \pi \ell$. This can be interpreted as stemming from a system with $kN$ charge carriers in the magnetic field $B$. The Hamiltonian for these $kN$ particules is the direct sum of $k$ copies of the Hamiltonian (\ref{totalhamiltonian}). In this new Hamiltonian, the charge carriers in differents copies do not interact. This means that the interaction potential is not exactly the same as the one used in (\ref{totalhamiltonian}). Nevertheless, in the computation of $\sigma_{xy}$ performed in (I), the \emph{explicit} expression for the potential was needed when calculating $\sigma_{xy}$: only some gap properties were assumed in order ensure the validity of the Kubo formula. Let us now make the natural physical hypothesis that the new potential has the properties needed for the Kubo formula to be used. Then, because the relation $kNB=2 \pi \ell$ holds, the use of formula (45) of (I), yields 
\begin{equation}
\sigma_{xy}(kN,B)=2 \frac{(kN)^2}{2\pi \ell}
\end{equation}
Now, the conductivity for $kN$ charge carriers is $k$ times the conductivity of $N$ charge carriers: $\sigma_{xy}(kN,B)=k \sigma_{xy}(N,B)$. Then one has
\begin{equation}
\label{sigmaxy}
\sigma_{xy}(N,B)=2 \frac{(kN)^2}{2\pi \ell}\frac{1}{k} = 2 \frac{N^2 k}{2\pi \ell} = 2\frac{N}{B}
\end{equation}
As expected, provided the physical hypothesis made on the potential are valid, the result coincides with the one obtained in (I).

\section{Physical discussion and conclusion}

\label{physicalcomments}
\setcounter{equation}{0}

In the previous section, we have shown that for all the rational values of $NB/2 \pi$, says $NB/2\pi = \ell/k$, the transverse conductivity $\sigma$ takes integers or fractionnal values while the wave functions in $\cD$ are rigidily linked with the topological and geometrical structures. The purpose of this section is to discuss heuristically what additionnal physical information could be extracted from these mathematical structures when $NB/2 \pi$ differs from a rational value, taking into account some plausible physical assumptions.

One of the main features of the real Quantum Hall experiments is the occurence of plateaux in the transverse conductivity when $B$ varies while no plateau can show up within the system described by fig.~\ref{fig-general} and (\ref{totalhamiltonian}) (plus boundary conditions) that we have analysed up to now, since $\sigma$ is linear in $B$ for all the rational values of $NB/2 \pi$. Consider the case of integer QHE. Then, it is believed that charge carriers are localized by disorder effects (which are basically induced by the presence of impurities in the Hall sample). This can be shown in Landau type models on $\gR^2$ in which the localized states modify the Hamiltonian spectrum. At least in a weak disorder regime, for which the typical energy of the disorder potential is small compared to the cyclotron energy, each Landau level is broadened by disorder giving rise to a Landau band associated with an extended state with energy centered around the corresponding Landau energy. Now, when $B$ varies in such a way that the system stays in such a band, the number of states involved in the charge transport remains constant so that the resulting conductivity is constant. 

A rigourous extension of the present analysis to take into account the disorder effects through the introduction of a suitable random potential in (\ref{totalhamiltonian}) is a task that goes beyond the scope of the present paper. In the following, we will not perform a mathematical analysis but instead we will propose a heuristic scenario which might correspond to the situation with non zero disorder, introducing for this purpose some plausible assumptions. The main point underlying our discussion has in fact been implicitely suggested in \cite{NiuThouWu:85}. In fact, these autors gave convincing arguments suggesting that in the presence of disorder\footnote{at least for weak disorder} the system retains (a sufficient amount of its) underlying geometrical structure so that the transverse conductivity remains quantized. Keeping this in mind, let us assume that the following two hypothesis are valid:
\begin{description}
\item[i)] The geometrical structures exhibited in the previous sections for rational values of $NB/2 \pi$ are not altered when a weak disorder potential is incorporated so that disorder effects would manifest themselves in the boundary conditions for the wave functions as phase changes compensating deviations of $NB/2 \pi$ from its rational values.
\item[ii)] The wave function $\widehat{\psi}(x,y,\gamma, \eta)$ for non zero disorder are related to those at zero disorder through a phase change $\xi(x,y)$ as
\begin{equation}
\label{phase}
\widehat{\psi}(x,y,\gamma, \eta) = e^{i \xi(x,y)} \psi(x,y, \gamma, \eta)
\end{equation}
which is suggested by scattering theory.
\end{description}
Now, in a weak disorder regime, the phase change $e^{i \xi(x,y)}$ in (\ref{phase}) can be expected to be close to unity,\footnote{see e.g. Prange in \cite{Ston:92} and ref. therein} $e^{i \xi(x,y)}\simeq 1$, so that $\xi(x,y)$ can be parametrized by $\xi(x,y) = 2 \pi r + \varepsilon(x,y)$ where the integer $r$ depends on the location $(x,y)$ and $\varepsilon(x,y)$ is bounded: $|\varepsilon(x,y)| \leq \delta/2$. In the spirit of hypothesis i), $\delta$ represents the upper value of the disorder contribution for which the geometrical structures exhibited in the previous sections are preserved giving rise to a quantized transverse conductivity. Then using (\ref{phase}), the boundary conditions (\ref{boundarypsitorus1})--(\ref{boundarypsitorus2}) on the wave functions relevant for the case $NB/2 \pi$ integer must be given by:
\begin{subequations}
\begin{align}
\label{boundarypsihattorus1}
\widehat{\psi}(x+1,y,\gamma,\eta) &= e^{i \gamma + \frac{i}{2} NB y} e^{i(\varepsilon(x+1,y) - \varepsilon(x,y))} \widehat{\psi}(x,y,\gamma,\eta) \\
\label{boundarypsihattorus2}
\widehat{\psi}(x,y+1,\gamma,\eta) &= e^{i \eta - \frac{i}{2} NB x} e^{i(\varepsilon(x,y+1) - \varepsilon(x,y))} \widehat{\psi}(x,y,\gamma,\eta)\ .
\end{align}
\end{subequations}
Then, $\widehat{\psi}$ can be extended in a unique way on the whole $\gR^2$ plane provided
\begin{equation}
e^{i(NB + \varepsilon(x,y+1) - \varepsilon(x+1,y))} = 1
\end{equation}
or
\begin{equation}
\label{extensioncondition}
NB + \varepsilon(x,y+1) - \varepsilon(x+1,y)) = 2\pi \ell
\end{equation}
Since one has $|\varepsilon(x,y)| \leq \delta/2$, (\ref{extensioncondition}) gives rise to the range of variations for $NB/2\pi$ around an integer value for which, according to hypothesis i), the transverse conductivity is quantized; namely, one obtains
\begin{equation}
\label{inequalityinteger}
|NB - 2\pi \ell| \leq \delta
\end{equation}
From a similar analysis applied to the boundary conditions relevant for the rational case, it can be easily realized that (\ref{inequalityinteger}) becomes 
\begin{equation}
\label{inequalityfraction}
\left|NB - 2\pi \frac{\ell}{k}\right| \leq \delta
\end{equation}
In view of (\ref{inequalityfraction}), the main assumptions i), if finally correct, leads to the physical conclusions that the transverse conductivity stays fixed to its quantized value (\ref{sigmaxy}) for $B$ varying in the range $\left[ -2 \pi \frac{\ell \delta}{N}, 2 \pi \frac{\ell \delta}{N} \right]$ for fixed $N$, which clearly can be viewed as a plateau.

The use of formula (\ref{inequalityinteger}) permits one to obtain some insight into the global behavior of the physical system studied in this paper when $B$ is varied and the contribution stemming from the disorder comes into play. Recall that $\nu=\frac{N}{N_\phi}$ where $N_\phi$ is the number of magnetic fluxes through the sample and that in our units, $\frac{1}{2\pi} = e^2/h$ so that $N_\phi = B/2\pi$. Let us assume that the number of electrons $N$ is kept fixed to some value, says $N=N_0$.\footnote{In the Quantum Hall experiment, this would be obtained by fixing the gate voltage controlling the electron density.} Then, from the analysis of section~\ref{rationnalcase}, we know that when $B$ reaches the value $B^\ast_{\ell/k} = \frac{2\pi}{N_0}\frac{\ell}{k}$, the transverse conductivity must be $\sigma^\ast_{\ell/k} = \frac{e^2}{h} \left(2 N_0^2 \frac{k}{\ell}\right) = \left( \frac{e^2}{h}\right) s_0$ for integer or fractional $s_0$. Now, according to the analysis presented at the beginning of this section, the occurence of disorder will not modify the value for the transverse conductivity, $\sigma = \sigma^\ast_{\ell/k}$, provided
\begin{equation}
\label{inequality}
\frac{2}{s_0} - \Delta \leq \frac{1}{\nu} \leq \frac{2}{s_0} + \Delta
\end{equation}
still holds, where $\Delta = \frac{e^2}{h} (\delta / N_0^2)$ and the expression for $\sigma^\ast_{\ell/k}$ together with the definition of the filling factor $\nu$ have been used. Then, one concludes from (\ref{inequality}) that each value for $s_0\in \gQ$ corresponds to a domain in the $\left( \frac{1}{\nu}, \Delta \right)$ plane whose associated state has transverse conductivity equal to $\sigma^\ast = (e^2/h) s_0$. Such a domain would therefore represent a phase in the $\left( \frac{1}{\nu}, \Delta \right)$ plane which could be fully characterized from the properties of its corresponding wave function and energy derived from the Hamiltonian (\ref{totalhamiltonian}) enlarged with a disorder potential. In particular, this would permits one to compare the respective energies of states corresponding to the (partially) overlapping stemming from (\ref{inequality}) in the $\left( \frac{1}{\nu}, \Delta \right)$ plane in order to determine which state is energically favored with respect to the other ones, therefore removing the spurious physically meaningless overlaps. Unfortunately, the full corresponding Schr\"odinger problem cannot be explicitely solved in the present case.

\begin{figure}
\begin{center}
\includegraphics{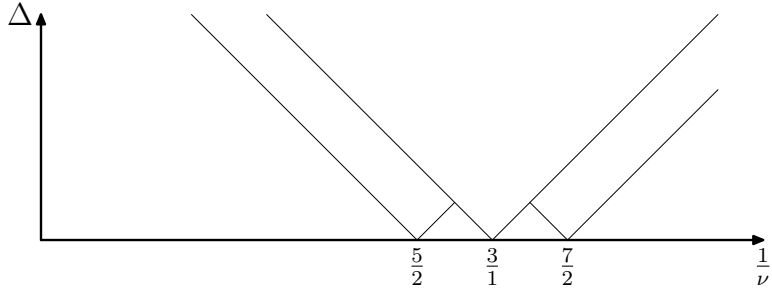}
\end{center}
\caption{Phase architecture in the $\left( \frac{1}{\nu}, \Delta \right)$ plane for the 3 states $s_0 = 1/3,\ s_0=2/5,\ s_0=2/7$.}
\label{fig-3phases}
\end{figure}

Besides, we further note that the value of the transverse conductivity we have found is twice the value which is expected in realistic Quantum Hall experiments. We have carefully checked that this factor 2 does appear when calculating \emph{in extenso} the conductivity through the Kubo formula\footnote{Usual assumptions on the validity of the Kubo formula are assumed here.} for the system depicted on fig.~\ref{fig-general} involving spinless charged particles and submitted to the boundary conditions (\ref{boundarypsi12}). Stricto sensu, realistic Quantum Hall systems involve spin carrying electrons in a strong magnetic field and are therefore somehow different than the idealized system considered here for which the magnitude of the magnetic field seen by the spinless charge carriers is not specified. It is concevable that the extension of the present study to the case of charge carriers with spin may lead, in some strong magnetic field limit, to some mechanism and/or degeneracy of the ground state that triggers the compensation of the above mentionned factor 2. If we assume that this finally happens, then eq.~(\ref{inequality}) now becomes
\begin{equation}
\label{inequality2}
\frac{1}{s_0} - \Delta \leq \frac{1}{\nu} \leq \frac{1}{s_0} + \Delta
\end{equation}
where $\Delta$ has been defined above and is associated with the disorder. Consider now the following odd denominator states $s_0 = 1/3,\ s_0=2/5,\ s_0=2/7$ for which the corresponding energies $E_{s_0}$ have been evaluated to be such that $E_{1/3} < E_{2/7} < E_{2/5}$ \cite{LHDM}. This, combined with (\ref{inequality2}) suggests the phase architecture in the $\left( \frac{1}{\nu}, \Delta \right)$ plane depicted on fig.~\ref{fig-3phases}. In this figure, we have used the plausible assumption that when two domains overlap, the indexing stable state corresponds to the one with lowest energy. It can be checked that successive actions of the Landau shift operator $\nu \mapsto \nu +1$ and of the flux attachment operator $\frac{1}{\nu} \mapsto \frac{1}{\nu} + 2$ generate a phase diagram similar to the one proposed in \cite{CGMW:02}.

Within the mathematical formalism introduced in (I), salient geometrical structures appear, depending on the values reached by the ratio $NB/2\pi$. These structures control possibly the (integral or fractionnal) quantization of the Hall conductivity. The simplest situation corresponds to integer values for $NB/2\pi$ and gives rise, as shown in (I), to a Hall conductivity taking integer or fractionnal values. In the present paper, we have analysed some of the main (new) geometrical features underlying the Hall system described in section~\ref{previousresults} in the more complicated situation where $NB/2\pi$ differs from an integer, paying special attention to the geometrical characterization of the space on which monovaluated wave functions can be defined. When $NB/2\pi$ takes a rational value $\ell/k$, this space is a Riemann surface that we have completely determined in term of the number of punctures and genus, both related to the denominator of the rational value. Provided the potential satisfies an additional condition, the explicit computation of the Kubo formula performed in (I) when $NB/2\pi$ is integer can be extended to rational values for $NB/2\pi$ and gives rise again to Hall conductivity taking integer or fractionnal values. When $NB/2\pi$ is irrationnal, we have found that the above space can be related to the graph of a free group with two generators. In that latter case however, the (explicit) computation of the Kubo formula cannot be performed. Finally, we have discussed from a phenomenological viewpoint what might be expected for a Hall system with $NB/2\pi$ differing slightly from a rational value.

We presented some physical arguments suggesting that interesting physical features (to be extracted) should be encoded  in our formalism. This suggests to examine more closely the irrationnal case and in particular to determine how the geometrical structures that occur in the rationnal case evolve when $NB/2\pi$ differs from rationnal values. Presumably, those ``deformations'' should be better treated within an algebraic framework. We notice that an indication in this direction are already encoded in the unitary symmetries that we have exhibited in (I). Recall that these latter connect rigidely (structures in) the direct space to (structures in) the reciprocical space. Besides, the two operators generated the unitary symmetries bear some similarity with the irrationnal rotation algebra, the $C^\ast$-algebra introduced in \cite{Rief:81, Conn:86}. As indicated in (I) and in the present paper, in the direct space quantization of  $NB/2\pi$ relies on the existence of some geometrical fiber bundle, while in the reciprocical space, the quantization of the Hall conductance stemms from some geometrical counterpart fiber bundle that is induced by the unitary symmetries. Now, from an algebraic viewpoint, this is strongly reminiscent of $K$-theory in direct space connected to $K$-theory in reciprocical space. But a natural framework to deal with both $K$-theories \emph{at the same time} is provided by the $KK$-theory of Kasparov (for a review see e.g. \cite{Skan:91}). The corresponding implications are presently under study.

\appendix

\section{Topological characterization of $\cS_k$}
\label{pi1}
\setcounter{equation}{0}

In this appendix, we characterize the Riemann surface with puncture $\cS_k$ introduced in section~\ref{rationnalcase}. The only missing information about this surface is its genus: from its construction, one knows that it is orientable, without boundary and with $k^2$ punctures. A elegant way to calculate its genus is to compute its first homotopy group $\pi_1(\cS_k)$. Indeed, using standard techniques in algebraic topology and the well known theorem of van~Kampen, one can easily rely the genus $g$, the number of punctures $t$ and the number of generators $n$ of the group $\pi_1(\cS_k)$ by the formula
\begin{equation}
n = 2g+t-1
\end{equation}
In order to get $g$, we only need to compute the number of generators of $\pi_1(\cS_k)$. This can be done using the van~Kampen theorem. Recall that this theorem permits one to compute the first homotopy group of a space $X$ using a decomposition $X = U \cup V$, where $U, V$ and $U \cap V$ are open, nonempty and arcwise connected, by the relation
\begin{equation*}
\pi_1(X) \simeq \pi_1(U) \star_{\pi_1(U\cap V)} \pi_1(V)
\end{equation*}
where $G_1 \star_H G_2$ is the free group amalgation of the group $G_1$ by $G_2$ over a group $H$ when there exists two morphisms $\phi_i : H \rightarrow G_i$. We refer to \cite{Bred:95} for the complete definition. We will make use of this formula when $\pi_1(U\cap V)$ is trivial, which means that the product is a free product. This means that the number of generators of $\pi_1(X)$ is the sum of the number of generators of $\pi_1(U)$ and $\pi_1(V)$.

\begin{figure}
\begin{center}
\includegraphics{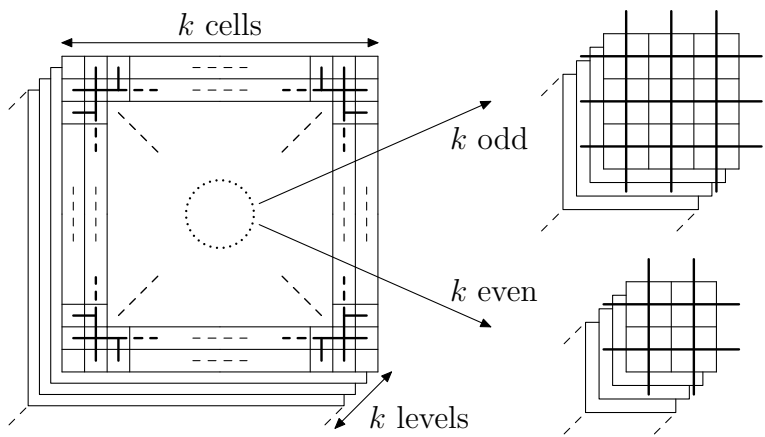}
\end{center}
\caption{General situation for generic $k$, with emphase for the last (inner) shell which depends on the parity of $k$.}
\label{fig-computation}
\end{figure}

The idea to compute $\pi_1(\cS_k)$ is to reconstruct the space $\cS_k$ by adding, step by step, the $k^3$ squares from which it is composed. The first space we take is a circle, composed of the $4k(k-1)$ peripherical squares of the $k^2$ stacks. As illustrated on fig.~\ref{fig-k2} and fig.~\ref{fig-k3}, the links connecting these squares form a global circle.  This property is easily demonstrated using the area rule. Let us now decompose these $k^2$ stacks of square into successive shells from outside to inside, as illustrated on fig.~\ref{fig-computation}. The shell labeled by $s=0$ is the one just described, composed of $4k(k-1)$ squares. The next one, labeled by $s=1$, is composed of $4(k-2-1)$ stacks, which gives $4k(k-2-1)$ squares. The shell labeled $s$ is composed of $4(k-2s-1)$ stacks, and $4k(k-2s-1)$ squares. The last internal shell depends on the parity of $k$: for $k$ odd it constains $1$ stack of $k$ squares, and for $k$ even, $4$ stacks. In the odd case, we will not consider the last shell as the one containing only $1$ stack, because it plays a different role in the following computation, but the one just before, containing $8$ stacks, which corresponds to $s=\frac{k-3}{2}$. In the even case, the last shell corresponds to $s=\frac{k-2}{2}$. This means also that we have to consider $k>3$ in the following analysis.

Let us now explain the reconstruction method. In the first step, we take $U$ to be the circle composed of the $4k(k-1)$ peripherical squares. First, we wish to add a square in the corner of the shell $s=1$. This square is connected to the shell $s=0$ by two links, as symbolised on fig.~\ref{fig-computation}. Notice that these two links do not connect the square to the same level in the stacks of the shell $s=0$. It is possible to close these links along the shell $s=0$, and then the space $V$ we take is just topologically a circle. When the four squares in the corner of the shell $s=1$ has been added, one can add the other squares one by one exactly by the same procedure. At each step the space $U$ is growing, and the space $V$ is topologically equivalent to a circle. We can repeat this procedure for all the shells. One has also to add the external links connected to the shell $s=0$, which are also topologically some circles.

Let us compute the number of generators of the first homotopy group from this construction using the van~Kampen theorem. At each step, the space $V$ is a circle, which means that $\pi_1(V)$ has one generator, and the intersection $U\cap V$ is contractible, which means that $\pi_1(U\cap V)$ is trivial. Then at each step, the free group amalgation consists to add exactly one generator to the homotopy group. In order to get the number of generators of $\pi_1(\cS_k)$, one has to compute the number of circles one has to add to reconstruct $\cS_k$ from the original circle. 
\begin{description}
\item[The shell $s=0$.] This shell is a circle, it contributes to $1$ generator. This is the first space $U$.

\item[The external links.] Each corner of the shell $s=0$ is connected to exactly two other corners: they contribute to $4k$ generators. Each other square is connected to exactly one other square: they contribute to $2k(k-2)$ generators.

\item[The generic shell $s$.] There are $4k$ corners which contribute each to $1$ generator. For the other squares, one has to connect then to the shell $s-1$ and to the corners. A carefull analysis shows that there are $4k(k-2s-1)$ circles to add to connect them.

\end{description}
Let us treat separetely the odd and even cases for the last shell.
\begin{description}
\item[The odd case $k=2p+1$.] One has to add one stack of $k$ squares to the shell $s=\frac{k-3}{2}=p-1$. Each square in this stack is connected to $4$ squares of the previous shell (see for instance the left graph of  fig.~\ref{fig-k3}), but one need only to add $3$ circles to connect such a square. So this shell contributes to $3k$ generators.

\item[The even case $k=2p$.] One has to connect $4$ stacks of $k$ squares. These squares are corners, so they contribute to $4k$ generators to connect them to the previous shell. These squares are linked together, this adds $4k$ circles. Notice that this last contribution is the same as the contribution of a generic shell with $s=\frac{k-2}{2}$: $4k(k-2\frac{k-2}{2}-1) = 4k$. This means that this contribution can be added to the generic one.
\end{description}
Let us collect all these contributions in the two cases:
\begin{description}
\item[The odd case $k=2p+1$.] 
\begin{equation*}
1 + 4k + 2k(k-2) + \sum_{s=1}^{p-1} \left( 4k + 4k(k-2s-1) \right) + 3k = 1 + k^3
\end{equation*}
\item[The even case $k=2p$.] 
\begin{equation*}
1 + 4k + 2k(k-2) + \sum_{s=1}^{p-1} \left( 4k + 4k(k-2s-1) \right)  = 1 + k^3
\end{equation*}
\end{description}
The result is independant of the parity of $k$.\footnote{This may mean that our demonstration is not the most elegant one!} Moreover, this computation supposes $k>3$, but the number of generators of $\pi_1(\cS_k)$ is also $1+k^3$ when $k=1,2,3$ as can be directly verified. So, the main result is that
\begin{center}
\emph{the number of generators of $\pi_1(\cS_k)$ is $1+k^3$.}
\end{center}
Knowing that $(\cS_k)$ has $k^2$ punctures, this implies that
\begin{center}
\emph{the number of handles of $\cS_k$ is $g_k=\frac{1}{2}(2 + k^2(k-1))$.}
\end{center}

\end{document}